\documentclass[twocolumn,amsmath,amssymb,prl,groupedaddress,superscriptaddress]{revtex4}
\usepackage{graphicx,color}
\usepackage{dcolumn}
\usepackage{bm}
\newcounter{one}
\usepackage{braket}
\usepackage[tight]{subfigure}

\begin{document}

\title{On-demand photonic entanglement synthesizer}

\author{Shuntaro Takeda}
\email{takeda@ap.t.u-tokyo.ac.jp}
\affiliation{Department of Applied Physics, School of Engineering, The University of Tokyo,\\ 7-3-1 Hongo, Bunkyo-ku, Tokyo 113-8656, Japan}
\affiliation{JST, PRESTO, 4-1-8 Honcho, Kawaguchi, Saitama, 332-0012, Japan}

\author{Kan Takase}
\affiliation{Department of Applied Physics, School of Engineering, The University of Tokyo,\\ 7-3-1 Hongo, Bunkyo-ku, Tokyo 113-8656, Japan}

\author{Akira Furusawa}
\email{akiraf@ap.t.u-tokyo.ac.jp}
\affiliation{Department of Applied Physics, School of Engineering, The University of Tokyo,\\ 7-3-1 Hongo, Bunkyo-ku, Tokyo 113-8656, Japan}

\date{\today}





\maketitle


{\bf
Quantum information protocols require various types of entanglement, such as 
Einstein-Podolsky-Rosen (EPR), Greenberger-Horne-Zeilinger (GHZ),
and cluster states~\cite{35Einstein,98Furusawa,03Bartlett,90Greenberger,99Hillery,00vanLoock,01Raussendorf,07vanLoock,06Menicucci}.
In optics, on-demand preparation of these states
has been realized by squeezed light sources~\cite{03Aoki,08Yukawa,13Yokoyama,14Chen},
but such experiments require different optical circuits for different entangled states, thus lacking versatility.
Here we demonstrate an on-demand entanglement synthesizer which programmably generates all these entangled states from a single squeezed light source. This is achieved by developing a loop-based circuit which is dynamically controllable at nanosecond timescale. We verify the generation of 5 different small-scale entangled states as well as a large-scale cluster state containing more than 1000 modes without changing the optical circuit itself.
Moreover, this circuit enables storage and release of one part of the generated entangled state, thus working as a quantum memory. 
This programmable loop-based circuit should open a way for a more general entanglement synthesizer~\cite{14Motes,07vanLoock}
and a scalable quantum processor~\cite{17Takeda}.
}

\begin{figure}[!b]
\begin{center}
\includegraphics[width=1\linewidth,clip]{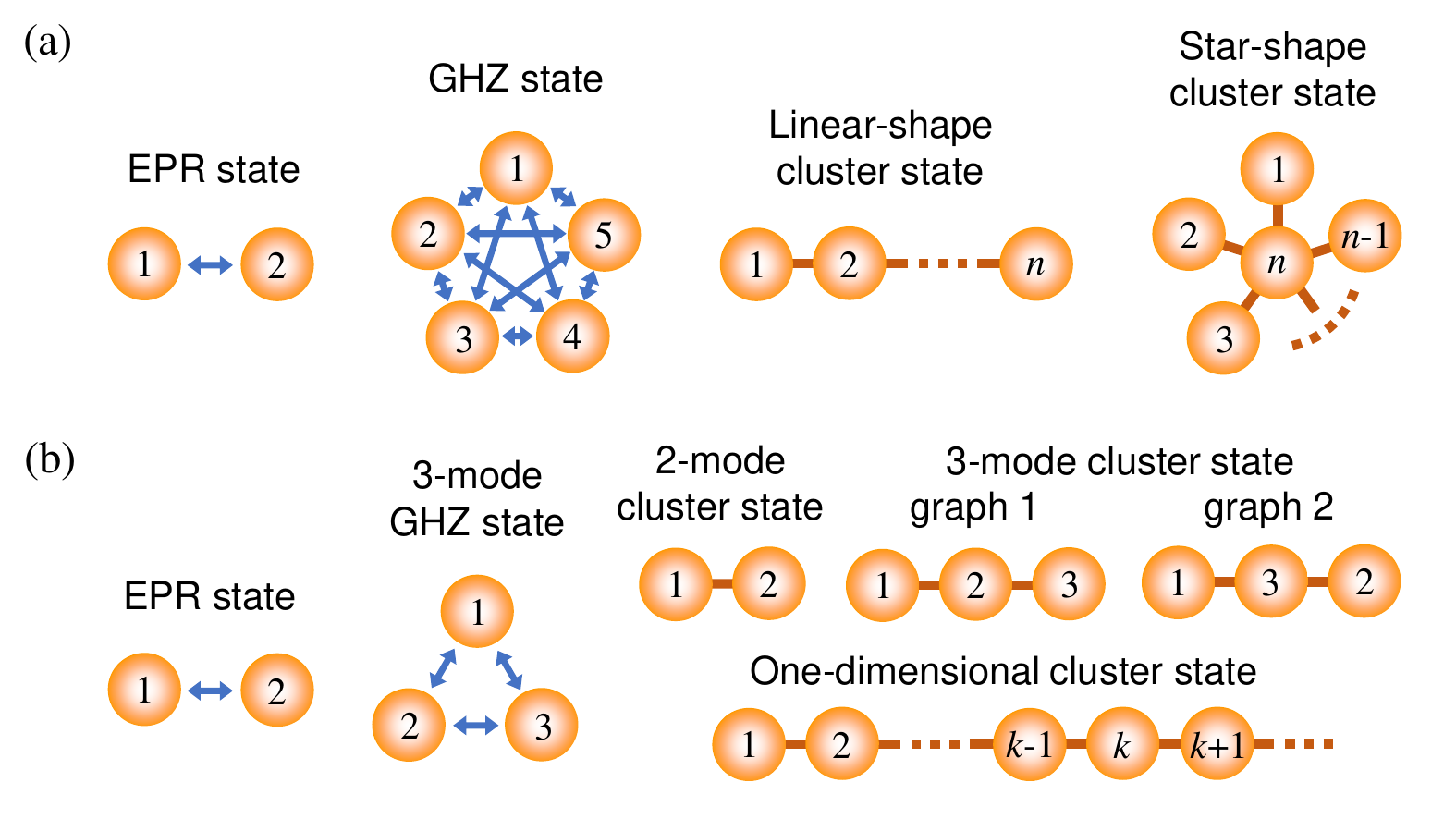}
\end{center}
\caption{
{\bf Various types of entanglement.}
(a) Types of entanglement which can be generated by our entanglement synthesizer.
(b) Types of entanglement which are actually generated and verified in this experiment. 
Orange spheres represent quantum modes.
Blue arrows connecting two modes mean that the connected nodes can communicate with each other by use of the entanglement.
Brown links connecting two modes mean that
an entangling gate to generate cluster states is applied between these modes.
}
\label{fig:TypesOfEntanglement}
\end{figure}

Entanglement is an essential resource for many quantum information protocols
in both qubit and continuous variable (CV) regimes.
However, different types of entanglement are required for different applications [Fig.~\ref{fig:TypesOfEntanglement}(a)].
The most commonly-used maximally-entangled state is a 2-mode EPR state~\cite{35Einstein},
which is the building block for two-party quantum communication and quantum logic gates based on quantum teleportation~\cite{98Furusawa,03Bartlett}.
Its generalized version is an $n$-mode GHZ state~\cite{90Greenberger,00vanLoock},
which is central to building a quantum network; 
this state, once shared between $n$ parties, enables any two of the $n$ parties to communicate with each other~\cite{99Hillery,00vanLoock}.
In terms of quantum computation, a special type of entanglement called cluster states
has attracted much attention as a universal resource for one-way quantum computation~\cite{{01Raussendorf},07vanLoock,06Menicucci}.

Thus far, the convenient and well-established method for deterministically preparing
photonic entangled state is to mix squeezed light via beam splitter networks
and generate entanglement in the CV regime~\cite{03Aoki, 08Yukawa, 13Yokoyama}.
By utilizing squeezed light sources multiplexed in time~\cite{13Yokoyama} or frequency~\cite{14Chen} domain,
generation of large-scale entangled states has also been demonstrated recently.
In these experiments, however, optical setups are designed to produce specific entangled states.
In other words, the optical setup has to be modified to produce different entangled states, thus lacking versatility.
A few experiments have reported programmable characterization of several types of entanglement in multimode quantum states
by post-processing on measurement results~\cite{12Armstrong} or changing measurement basis~\cite{14Roslund,17Cai}.
However, directly synthesizing various entangled states in a programmable and deterministic way is still a challenging task.

\begin{figure}[!t]
\begin{center}
\includegraphics[width=1\linewidth,clip]{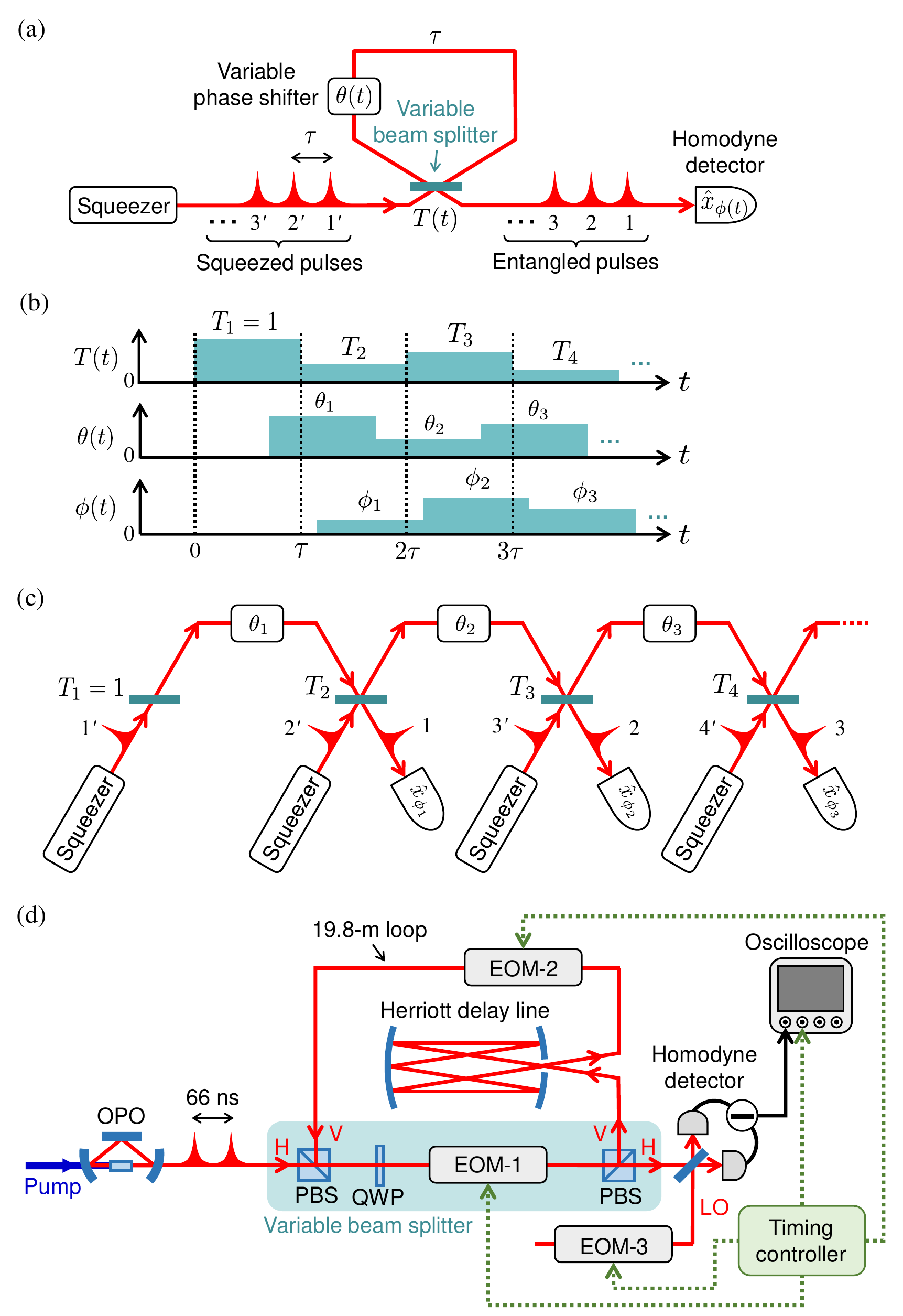}
\end{center}
\caption{
{\bf Schematic of an on-demand entanglement synthesizer.}
(a) Conceptual schematic.
(b) Time sequence for changing system parameters.
(c) Equivalent circuit.
(d) Experimental setup. See Methods for details.
``H'' and ``V'' denote horizontal and vertical polarization, respectively.
OPO, optical parametric oscillator;
PBS, polarizing beam splitter;
QWP, quarter wave plate;
EOM, electro-optic modulator;
LO, local oscillator. }
\label{fig:Schematic}
\end{figure}

Here we propose an on-demand photonic entanglement synthesizer
which can programmably produce an important set of entangled states,
including an EPR state, an $n$-mode GHZ state,
and an $n$-mode linear- or star-shape cluster state for any $n\ge2$ [Fig.~\ref{fig:TypesOfEntanglement}(a)].
The conceptual schematic of the synthesizer is shown in Fig.~\ref{fig:Schematic}(a).
Squeezed optical pulses are sequentially produced from a single squeezer,
and injected into a loop circuit whose round-trip time $\tau$ is equivalent to the time interval between the pulses.
This loop includes a beam splitter with variable transmissivity $T(t)$ and a phase shifter with variable phase shift $\theta(t)$,
where $t$ denotes time.
After transmitting the loop, the pulses are sent to a homodyne detector with a tunable measurement basis
$\hat{x}_{\phi(t)}=\hat{x}\cos\phi(t)+\hat{p}\sin\phi(t)$, where $\hat{x}$ and $\hat{p}$ are quadrature operators.
By dynamically changing $T(t)$, $\theta(t)$, and $\phi(t)$ for each pulse as in Fig.~\ref{fig:Schematic}(b),
this circuit can synthesize various entangled states from the squeezed pulses and analyze them.
This functionality can be understood by considering an equivalent circuit in Fig.~\ref{fig:Schematic}(c).
Here, the conversion from the squeezed pulses $1^\prime, 2^\prime, \cdots$ to the output pulses $1, 2, \cdots$ in Fig.~\ref{fig:Schematic}(c)
is completely equivalent to the corresponding conversion in Fig.~\ref{fig:Schematic}(a).
It is known that all of the entangled states in Fig.~\ref{fig:TypesOfEntanglement}(a) can be produced in the circuit of Fig.~\ref{fig:Schematic}(c)
as long as the beam splitter transmissivity $(T_1, T_2, \cdots)$ and phase shift $(\theta_1, \theta_2,\cdots)$ are arbitrarily tunable~\cite{00vanLoock,15Ukai} (see Methods).
However, the circuit in Fig.~\ref{fig:Schematic}(c) lacks scalability since one additional entangled mode requires one additional squeezer, beam splitter and detector.
In contrast, our loop-based synthesizer in Fig.~\ref{fig:Schematic}(a) dramatically decreases the complexity of the optical circuit,
and even more, can produce any of these entangled states by appropriately programming
the control sequence in Fig.~\ref{fig:Schematic}(b), without changing the optical circuit itself.

\begin{table*}[ht]
\begin{tabular}{|c|c|c|c|c|} \hline
Type of entanglement & $(T_1, T_2, \cdots)$ & $(\theta_1, \theta_2, \cdots)$ & Inseparability parameter & Measured value \\ \hline
EPR state & $(1, \frac12, 1)$ & $(90^\circ, 0^\circ)$ & $\braket{[\Delta(\hat{x}_1-\hat{x}_2)]^2}+\braket{[\Delta(\hat{p}_1+\hat{p}_2)]^2}$ & $0.44\pm0.01$ \\ \hline
3-mode GHZ state & $(1, \frac13, \frac12, 1)$ & $(90^\circ, 180^\circ, 0^\circ)$ &  $\braket{[\Delta(\hat{x}_1-\hat{x}_2)]^2}+\braket{[\Delta(\hat{p}_1+\hat{p}_2+\hat{p}_3)]^2}$ &  $0.65\pm0.01$  \\
& & & $\braket{[\Delta(\hat{x}_2-\hat{x}_3)]^2}+\braket{[\Delta(\hat{p}_1+\hat{p}_2+\hat{p}_3)]^2}$ &  $0.67\pm0.01$ \\
& & & $\braket{[\Delta(\hat{x}_1-\hat{x}_3)]^2}+\braket{[\Delta(\hat{p}_1+\hat{p}_2+\hat{p}_3)]^2}$ &  $0.70\pm0.01$ \\ \hline
2-mode cluster state & $(1, \frac12, 1)$ & $(90^\circ, 90^\circ)$ & $\braket{[\Delta(\hat{p}_1-\hat{x}_2)]^2}+\braket{[\Delta(\hat{p}_2-\hat{x}_1)]^2}$ &  $0.42\pm0.01$ \\ \hline
3-mode cluster  state & $(1, \frac23, \frac12, 1)$ & $(90^\circ, 90^\circ, 90^\circ)$ &  $\braket{[\Delta(\hat{p}_1-\hat{x}_2)]^2}+\braket{[\Delta(\hat{p}_2-\hat{x}_1-\hat{x}_3)]^2}$ &  $0.56\pm0.01$ \\
(graph 1)& & & $\braket{[\Delta(\hat{p}_3-\hat{x}_2)]^2}+\braket{[\Delta(\hat{p}_2-\hat{x}_1-\hat{x}_3)]^2}$ &  $0.54\pm0.01$ \\
& & & $\braket{[\Delta(\hat{p}_1-\hat{p}_3)]^2}+\braket{[\Delta(\hat{p}_2-\hat{x}_1-\hat{x}_3)]^2}$ &  $0.60\pm0.01$ \\ \hline
3-mode cluster  state & $(1, \frac13, \frac12, 1)$ & $(90^\circ, 180^\circ, 90^\circ)$ &  $\braket{[\Delta(\hat{p}_1-\hat{x}_3)]^2}+\braket{[\Delta(\hat{p}_3-\hat{x}_1-\hat{x}_2)]^2}$ &  $0.69\pm0.01$ \\
(graph 2)& & & $\braket{[\Delta(\hat{p}_2-\hat{x}_3)]^2}+\braket{[\Delta(\hat{p}_3-\hat{x}_1-\hat{x}_2)]^2}$ &  $0.65\pm0.01$ \\
& & & $\braket{[\Delta(\hat{p}_1-\hat{p}_2)]^2}+\braket{[\Delta(\hat{p}_3-\hat{x}_1-\hat{x}_2)]^2}$ &  $0.63\pm0.01$ \\ \hline
\end{tabular}
\caption{
{\bf Control sequence and inseparability parameters for various entangled states.}
$T(t)$ and $\theta(t)$ are controlled by the sequence in Fig.~\ref{fig:Schematic}(b)
with the setting values $(T_1, T_2, \cdots)$ and $(\theta_1, \theta_2, \cdots)$ defined in this table.
$\phi(t)$ is also controlled to measure the inseparability parameter for each state.
The generated state is inseparable when each inseparability parameter is below 1 ($\hbar=1/2$).
The expression of inseparability parameters are given in Refs.~\cite{00Duan,03vanLoock,08Yukawa,12Armstrong}.}
\label{tb:various_ent}
\end{table*}

This programmable loop-based circuit is, in fact, a core circuit to build more general photonic circuits.
By embedding this loop circuit in another large loop,
we can realize an arbitrary beam splitter network to combine input squeezed pulses~\cite{14Motes},
thereby synthesizing more general entangled states including an arbitrary cluster state~\cite{07vanLoock}.
Moreover, this circuit can be further extended to a universal quantum computer
by incorporating a programmable displacement operation based on the homodyne detector's signal
and another non-Gaussian light source~\cite{17Takeda}.
In these schemes, fault-tolerant quantum computation is possible even with finite level of squeezing~\cite{14Menicucci,17Takeda}.

We implement this synthesizer by a setup shown in Fig.~\ref{fig:Schematic}(d).
Here, squeezed optical pulses arrive at a 19.8-m loop every $\tau=66$ ns .
We develop a technique to dynamically change the beam splitter transmissivity, phase shift, and measurement basis within 20 ns,
and time-synchronize the switching of all these parameters at nanosecond timescale (See Methods).
As a demonstration of programmable entanglement generation,
we first program the synthesizer to generate 5 different small-scale entangled states,
including  an EPR state, a 3-mode GHZ state, a 2-mode cluster state,
and two 3-mode cluster states with different graphs,
as shown in Fig.~\ref{fig:TypesOfEntanglement}(b)
(graph 1 and 2 correspond to the linear and star shape, respectively; see Fig.~\ref{fig:TypesOfEntanglement}(a)).
In order to verify generation of the desired entanglement,
we measure the inseparability parameter (correlations of quadratures $\hat{x}_k$, $\hat{p}_k$, where $k$ is the mode number)
to evaluate inseparability criteria~\cite{00Duan,03vanLoock,08Yukawa,12Armstrong}.
The inseparability parameter below 1 ($\hbar=1/2$) is a sufficient condition
for the state to be fully inseparable.
Table~\ref{tb:various_ent} summarizes the control sequence of the system parameters
as well as the expression and measured values of the inseparability parameter for each state.
We see that all the values satisfy the inseparability criteria
and clearly demonstrate the programmable generation of 5 different entangled states.
Note that the current experimental setup is unable to synthesize more-than-3-mode GHZ and cluster states
(except for the large-scale cluster state described in the next paragraph)
for technical reasons, which can be overcome with a slight modification (See Methods).

\begin{figure}[!b]
\begin{center}
\includegraphics[width=1\linewidth,clip]{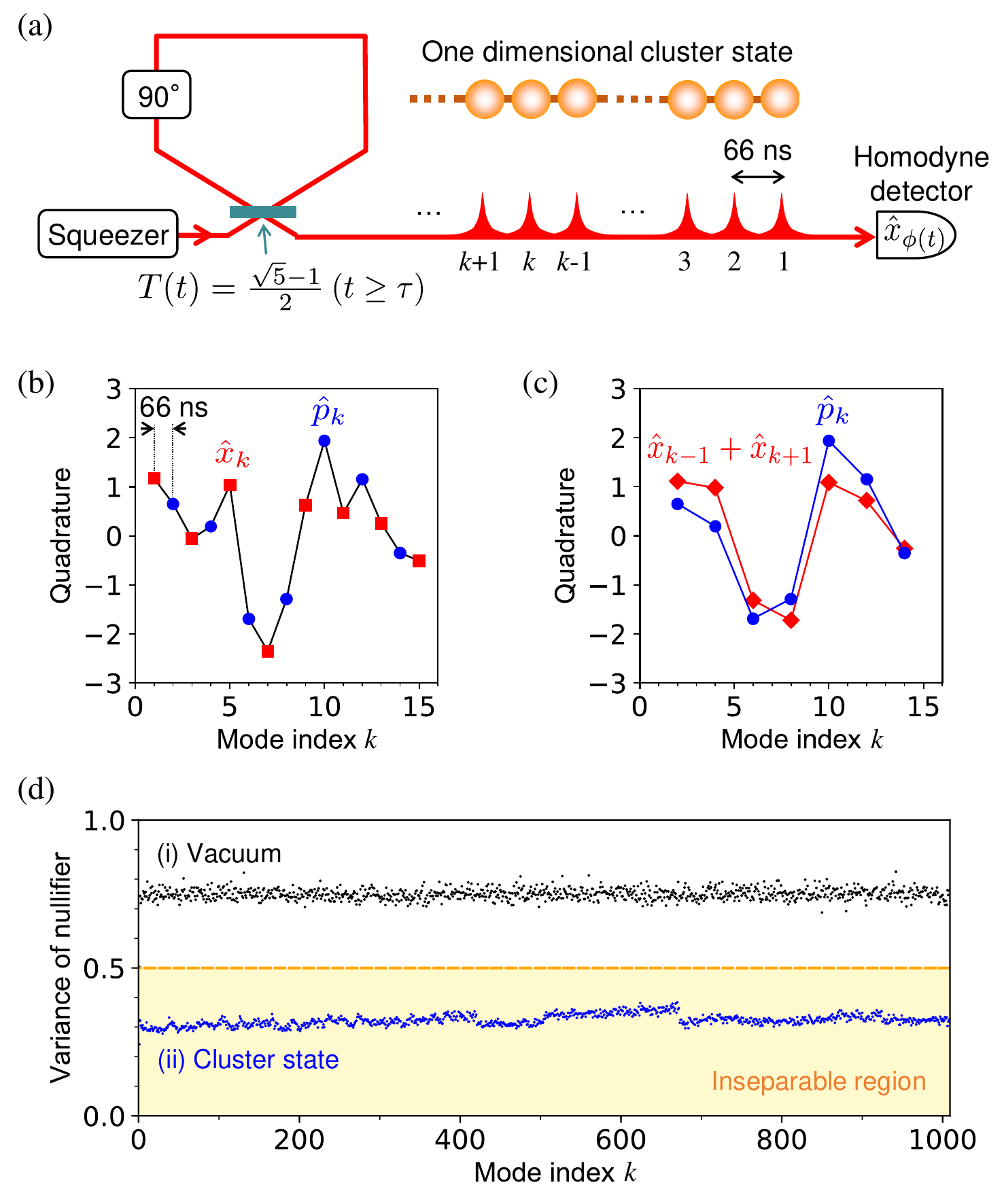}
\end{center}
\caption{
{\bf Generation of a one-dimensional cluster state.}
(a) Schematic.
(b) Single-shot measurement of quadratures for the first 15 modes.
$\hat{x}_k$ ($\hat{p}_k$) is measured for odd (even) number modes
and plotted as red squares (blue circles).
(c) Comparison between $\hat{p}_k$ (blue circles)
and $\hat{x}_{k-1}+\hat{x}_{k+1}$ (red diamonds).
(d) Measured variance of nullifier $\braket{\hat{\delta}_k^2}$
for (i) vacuum states (as a reference, black dots) and (ii) the cluster states (blue dots).
Standard error of each variance is around 0.01 and always below 0.03.
Yellow shaded area represents the inseparable region.
}
\label{fig:1Dcluster}
\end{figure}

Our entanglement synthesizer is not limited to producing small-scale entangled states,
but can produce a large-scale entangled state and thus possesses high scalability.
We demonstrate this scalability by generating a large-scale one-dimensional cluster state
[Fig.~\ref{fig:TypesOfEntanglement}(b)],
which is known to be a universal resource for single-mode one-way quantum computation for CVs~\cite{06Menicucci}.
This state can be produced by dynamically controlling the system parameters as
$T_1=1$, $T_k=(\sqrt{5}-1)/2$ ($k\ge2$), and $\theta_k=90^\circ$ ($k\ge1$).
Under this condition, a one-dimensional cluster state is continuously produced,
as shown in Fig.~\ref{fig:1Dcluster}(a) (see Methods).
This circuit is effectively equivalent to the cluster state generation proposed in Ref.~\cite{10Menicucci}.
The generated state can be characterized by a nullifier $\hat{\delta}_k$, defined as
\begin{align}
\hat{\delta}_k=
\begin{cases}
\hat{p}_1-\hat{x}_2 & (k=1) \\
\hat{p}_k-\hat{x}_{k-1}-\hat{x}_{k+1} & (k\ge2)
\end{cases}
\end{align}
and $\braket{\hat\delta_k^2}\to0$ in the limit of infinite squeezing.
The sufficient condition for the state to be inseparable is $\braket{\hat\delta_k^2}<1/2$ for all $k$~\cite{12Armstrong,13Yokoyama,08Yukawa}.

Figure~\ref{fig:1Dcluster}(b) shows the quadratures for the first 15 modes
acquired by setting the default measurement basis to $\hat{x}_k$ and
switching the basis to $\hat{p}_k$ only when $k$ is even.
The quadrature value looks randomly distributed, but once $\hat{x}_{k-1}+\hat{x}_{k+1}$ is calculated
and plotted with $\hat{p}_k$ as in Fig.~\ref{fig:1Dcluster}(c),
the correlation between these two values can be clearly observed.
This correlation results in the reduction of $\braket{\hat\delta_k^2}$ below $1/2$ in Fig.~\ref{fig:1Dcluster}(d),
which demonstrates the generation of the one-dimensional cluster state of more than 1000 modes.
Due to technical limitations associated with our control sequence, measurement time, and stability of the setup,
we stop the measurement at $k=1008$.
In principle, there is no theoretical limitation for the number of entangled modes in this method.

\begin{figure}[!t]
\begin{center}
\includegraphics[width=1\linewidth,clip]{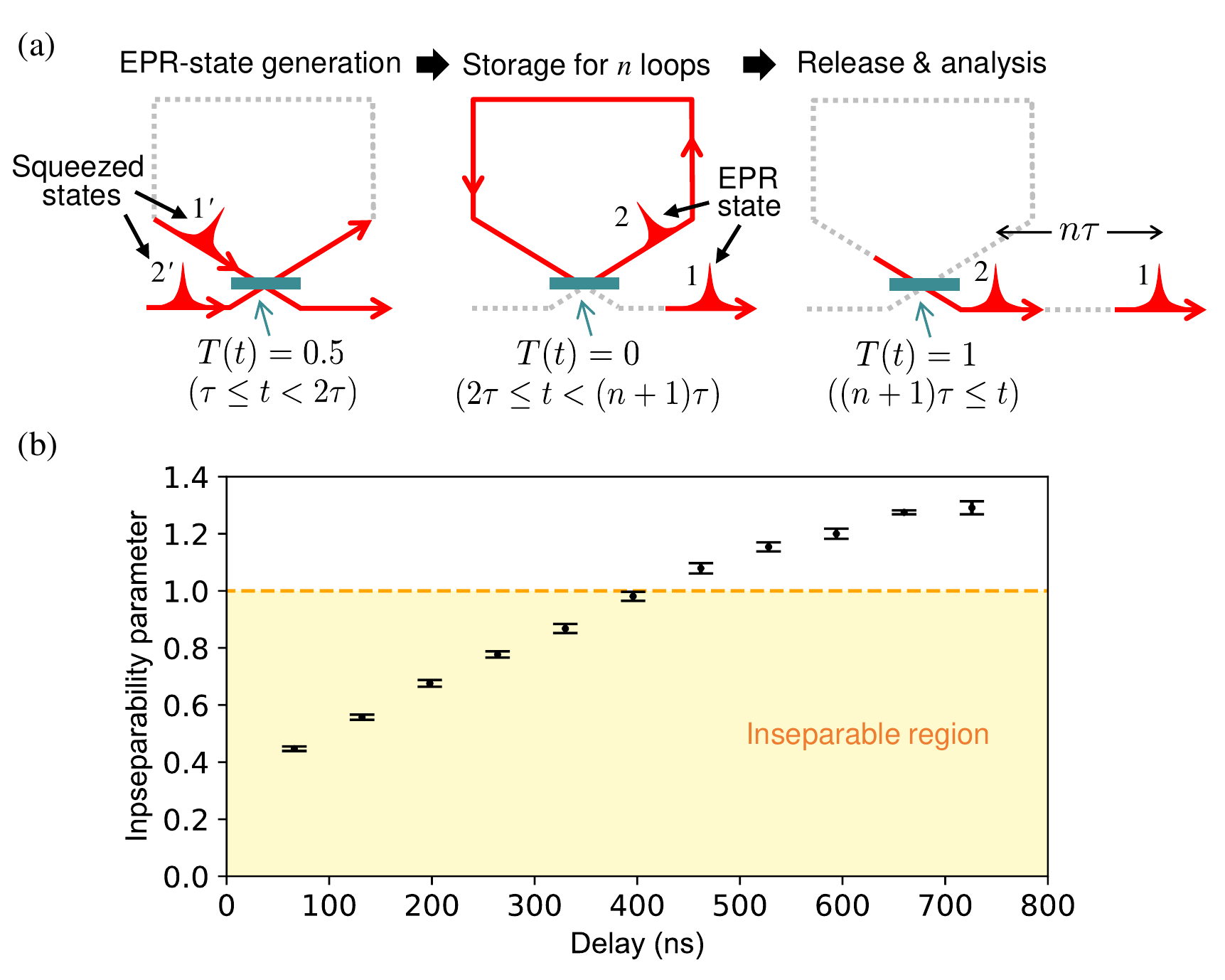}
\end{center}
\caption{
{\bf Storage of one part of an EPR state in the loop.}
(a) Control sequence.
(b) Measured inseparability parameter $\braket{[\Delta(\hat{x}_1-\hat{x}_2)]^2}+\braket{[\Delta(\hat{p}_1+\hat{p}_2)]^2}$
with standard error is plotted for each delay $n\tau$ ($\tau=66$ ns, $n=1, 2, \cdots, 11$).
Yellow shaded area represents the inseparable region.
}
\label{fig:Memory}
\end{figure}

The programmable loop circuit further allows us to
confine an optical pulse in the loop by keeping $T(t)=0$ and release it after $n$ loops, effectively acting as a quantum memory.
The ability to add a tunable delay to non-classical CV states plays a key role
for time synchronization in various quantum protocols~\cite{01Gottesman,03Bartlett,10Brask},
but there have been only a few memory experiments for CV entanglement~\cite{09Marino,11Jensen}.
In fact, a loop-based quantum memory
is a simple and versatile memory which limits neither the wavelength nor quantum state of light,
but it has been demonstrated only for single photons~\cite{02Pittman,15Kaneda}.
Here we demonstrate this functionality by first generating an EPR state in the loop,
then storing one part of the EPR state for $n$ loops, and finally releasing it [Fig.~\ref{fig:Memory}(a)].
In this scheme, one part of the EPR state is stored whereas the other is left propagating,
which is exactly the same situation as in quantum repeater protocol.
We measure the inseparability parameter for the EPR state after introducing the delay $n\tau$.
As shown in Fig.~\ref{fig:Memory}(b),
the inseparability parameter is below 1 and clearly satisfies the inseparability criterion up to $\sim400$ ns ($n=6$),
although it gradually degrades as the delay increases.
A theoretical simulation shows that
the lifetime of the EPR correlation in our system is dominantly limited by the phase fluctuation of $\sim7^\circ$ in the loop,
rather than the round-trip loss of $\sim7\%$;
a small phase drift in the loop is accumulated when the stored pulse circulates in the loop,
finally destroying the phase relation between the EPR pulses.
Therefore, the lifetime of our memory can be increased by improving the mechanical stability of the loop
or the feedback system to stabilize the phase.
Our loop-based memory can store any CV quantum states, such as non-Gaussian states,
by changing our squeezer to other quantum light sources~\cite{13Yukawa}.

In conclusion,
we have programmably generated and verified a variety of small- and large-scale entangled states
by dynamically controlling the beam splitter transmissivity, phase shift, and measurement basis
of a loop-based optical circuit at nanosecond timescale.
We have also demonstrated that this circuit can
work as a quantum memory by storing one part of an EPR state in the loop.
Our loop-based system is programmable and highly scalable,
offering a unique and versatile tool for future photonic quantum technologies.

\section{Methods}

\subsection{Experimental setup and data analysis}
We use a continuous-wave Ti:Sapphire laser at 860 nm.
Our optical parametric oscillator (OPO; the same design as in Ref.~\cite{17Shiozawa}) produces a squeezed light
with $\sim5$ dB of squeezing and $\sim8$ dB of anti-squeezing at low frequencies.
Our loop is built by a Herriott-type optical delay line~\cite{64Herriott} and has a round-trip length of 19.8 m.
Considering this loop length, we artificially divide the squeezed light
into 66-ns time bins. Each time bin is further divided into 20-ns switching time used for changing system parameters
and 46-ns processing time within which a squeezed optical pulse is defined.
The loop includes a variable beam splitter composed of two polarizing beam splitters (PBSs)
and one bulk-type polarization-rotation electro-optic modulator (EOM).
By inserting a quarter wave plate (QWP) between the PBSs, the transmissivity of the beam splitter is initially
set to $0.5$, and the EOM changes the transmissivity when it is triggered.
The variable phase shifter is realized by a bulk-type phase-modulation EOM,
shifting the phase from the initially locked value of  $0^\circ$ when it is triggered.
Finally, the pulses after the loop are mixed with a local oscillator (LO) beam and measured by a homodyne detector.
Here, a waveguide-type EOM in the LO's path can shift the phase $\phi(t)$ and thereby change the measurement basis $\hat{x}_{\phi(t)}$.
During the measurement, we periodically switch between two different settings at a rate of 2 kHz;
one is the feedback setting when the cavity length of the OPO and the relative phases of beams are actively locked by weakly-injected reference beams,
and the other is the measurement setting when the control sequence in Fig.~\ref{fig:Schematic}(b) is triggered and the data is acquired without the reference beams~\cite{13Yokoyama}.

In order to analyze the generated states,
we acquire the homodyne detector's signal by an oscilloscope at a sampling rate of 1.25 GHz.
5000 data frames are recorded to estimate each inseparability parameter and nullifier.
The quadrature of the $k$-th mode is extracted
by applying a temporal mode function $f_k(t)$ to each data frame, defined as~\cite{16Yoshikwa}
\begin{align}
f_k(t)\propto
\begin{cases}
e^{-\gamma^2(t-t_k)^2}(t-t_k) & (2|t-t_k|\le T) \\
0 & (\text{otherwise}),
\end{cases}
\end{align}
and normalized to be $\int_{-\infty}^{\infty}\left|f_k(t)\right|^2dt=1$.
The parameters used in this experiment are
$T=46$ ns, $\gamma=6\times10^7$ /s,
and $t_k=t_0+(k-1)\tau$, where $t_0$ is the optimized center position of the first mode and $\tau=66$ ns
is the interval between the modes.
Using these parameters,
we check the orthogonality of the neighboring modes
by applying $f_k(t)$ to the data frames for the shot noise signal
and confirming that the quadrature correlation between different modes are negligible~\cite{16Yoshikwa}.

\subsection{Working principle of variable beam splitter and phase shifter}

The EOMs for the variable beam splitter and phase shifter
contain a crystal of Rubidium Titanyl Phosphate
which is sandwiched between two electrodes.
By using fast high-voltage switches,
we can selectively apply 0 or $V_1$ volt to one of these electrodes
and 0 or $-V_2$ volt to the other electrode,
where $V_1>0$ and $V_2>0$ can be arbitrarily chosen in advance.
The net voltage applied to the crystal
can thus be switched among 0, $V_1$, $V_2$, $V_1+V_2$,
and these voltages determine the possible values of $T(t)$ and $\theta(t)$.
The rise/fall time for the switching is $\sim10$ ns.
In this system, it is not possible to switch $T(t)$ and $\theta(t)$ among more than 3 different target values in general.
Due to this limitation,
our setup is unable to generate GHZ or cluster states of more than 3 modes,
which require switching of $T(t)$ among 4 or more different values.
This limitation can be overcome by
modifying the EOM driving circuits or cascading more than one EOMs.

In the following, we introduce theoretical description of
the action of the variable beam splitter and phase shifter.
In Fig.~\ref{fig:Schematic}(c),
the $k$-th beam splitter with transmissivity $T_k$ ($k\ge2$) mixes
one mode from a squeezer (annihilation operator $\hat{a}_k^\prime=\hat{x}^\prime_{k}+i\hat{p}^\prime_{k}$)
and the other mode coming from the $(k-1)$-th beam splitter ($\hat{a}_{k-1}^{\prime\prime}$).
After this operation,
one of the output modes is measured ($\hat{a}_{k-1}=\hat{x}_{k-1}+i\hat{p}_{k-1}$),
while the other output mode becomes the input mode of the $(k+1)$-th beam splitter
after the phase shift of $\theta_k$ ($\hat{a}_{k}^{\prime\prime}$).
In Fig.~\ref{fig:Schematic}(d), the same operation is performed
with the variable beam splitter and variable phase shifter.
In the variable beam splitter,
the QWP initially introduces a relative-phase offset of $90^\circ$ between two diagonal polarizations,
thereby setting the default transmissivity to 0.5.
The polarization-rotation EOM introduces
an additional relative-phase shift of $2\delta_k\ge0$, which is proportional to the applied voltage.
Under this condition, the function of the $k$-th beam splitter and phase shifter in Fig.~\ref{fig:Schematic}(c)
is realized in Fig.~\ref{fig:Schematic}(d) as
\begin{align}
&\begin{pmatrix}
\hat{a}_{k-1} \\ \hat{a}_{k}^{\prime\prime}
\end{pmatrix} \nonumber \\
&=
\begin{pmatrix}
1 & 0 \\
0 & e^{i\theta_k}
\end{pmatrix}
\begin{pmatrix}
\sin(\delta_k+45^\circ) & -\cos(\delta_k+45^\circ) \\
\cos(\delta_k+45^\circ) & \sin(\delta_k+45^\circ)
\end{pmatrix}
\begin{pmatrix}
\hat{a}^{\prime\prime}_{k-1} \\ \hat{a}^\prime_{k}
\end{pmatrix}.
\label{eq:VBS}
\end{align}

The transmissivity of the variable beam splitter is defined by $T_k=\sin^2(\delta_k+45^\circ)$ in Eq.~(\ref{eq:VBS}).
By gradually increasing the applied voltage and thereby increasing $\delta_k$ from $0^\circ$ to $45^\circ$,
$T_k$ can be increased from 0.5 to 1. Thus any transmissivity between 0.5 and 1 can be chosen in this way.
When the transmissivity between 0 and 0.5 is required,
the voltage has to be further increased to set $\delta_k$ between $90^\circ$ and $135^\circ$.
In this region, however, the sign of the off-diagonal terms in Eq.~(\ref{eq:VBS}) flips.
This sign flip corresponds to the additional phase shift of $180^\circ$ before and after the beam splitter operation,
\begin{align}
&\begin{pmatrix}
\sqrt{T_k} & \sqrt{1-T_k}\\
-\sqrt{1-T_k} & \sqrt{T_k}
\end{pmatrix}\nonumber\\
&=
\begin{pmatrix}
1 & 0\\
0 & -1
\end{pmatrix}
\begin{pmatrix}
\sqrt{T_k} & -\sqrt{1-T_k}\\
\sqrt{1-T_k} & \sqrt{T_k}
\end{pmatrix}
\begin{pmatrix}
1 & 0\\
0 & -1
\end{pmatrix}.
\label{eq:VBS_flip}
\end{align}

\subsection{Generation of GHZ and star-shape cluster states}

It is known that an $n$-mode GHZ state ($n\ge2$, the case of $n=2$ corresponds to an EPR state) can be generated in the setup of Fig.~\ref{fig:Schematic}(c)
by setting
$T_1=T_{n+1}=1$, $T_k=1/(n-k+2)$ ($2\le k\le n$),
and $\theta_1=90^\circ$, $\theta_k=0$ ($2\le k\le n$)~\cite{00vanLoock}.
When all input modes are infinitely $\hat{x}$-squeezed vacuum states
(the input quadratures satisfy $\hat{x}^\prime_k=0$ for all $k$),
the quadratures of the output modes in this setting show
the correlation of the GHZ state,
\begin{align}
\hat{x}_k-\hat{x}_n= 0 \quad (1\le k\le n-1), \quad \sum_{k=1}^{n}\hat{p}_k= 0.
\label{eq:nullifier_GHZ}
\end{align}
Here we show that
the difference between the $n$-mode GHZ state and the $n$-mode star-shaped cluster state
is only local phase shifts.
We now replace $\hat{x}_l\to\hat{p}_l$ and $\hat{p}_l\to-\hat{x}_l$ for all $l$ in Eq.~(\ref{eq:nullifier_GHZ})
by redefining the quadratures.
We then introduce an additional phase rotation of $\theta_n=90^\circ$
to undo this replacement only for the $n$-th mode.
After these operations, Eq.~(\ref{eq:nullifier_GHZ}) transforms into 
\begin{align}
\hat{p}_k-\hat{x}_n=0\quad (1\le k\le n-1),\quad \hat{p}_n-\sum_{k=1}^{n-1}\hat{x}_k=0,
\end{align}
which are the definition of the $n$-mode star-shaped cluster state in Fig.~\ref{fig:TypesOfEntanglement}(a).
Therefore, the actual difference of the settings
for generating GHZ and star-shaped cluster states is only the value of $\theta_n$.

In this experiment, these settings are used for generating
the EPR state, 3-mode GHZ state, 2-mode cluster state, and 3-mode cluster state (graph 2).
When we generate 3-mode GHZ and cluster states (graph 2),
additional phase shifts of $180^\circ$ before and after the beam splitter
are introduced by the variable beam splitter with $T_2=1/3$,
as explained in Eq.~(\ref{eq:VBS_flip}).
The $180^\circ$ phase shift before the beam splitter has no effect
since it is applied to a squeezed vacuum state with $180^\circ$ rotational symmetry,
and the phase shift after the beam splitter
is cancelled out by setting $\theta_2=180^\circ$,
as shown in Table~\ref{tb:various_ent}.

\subsection{Generation of linear-shape cluster states}

The setup of Fig.~\ref{fig:Schematic}(c) can also produce
an $n$-mode linear-shape cluster state by setting
$T_1=T_{n+1}=1$, $T_k=F_{n-k+2}/F_{n-k+3}$ ($2\le k\le n$),
and $\theta_k=90^\circ$ ($1\le k\le n$)~\cite{15Ukai}.
Here, $F_k$ is a Fibonacci number defined by
$F_0=0$, $F_1=1$, $F_k=F_{k-1}+F_{k-2}$ ($k\ge2$)
and given by
\begin{align}
F_k=\frac{1}{\sqrt{5}}\left[\left(\frac{1+\sqrt5}{2}\right)^k-\left(\frac{1-\sqrt{5}}{2}\right)^k\right].
\end{align}
In this setting,
the $k$-th beam splitter with $T_k=F_{n-k+2}/F_{n-k+3}$, followed by the $k$-th phase shifter with $\theta_{k}=90^\circ$,
transforms the annihilation operators as
\begin{align}
&
\begin{pmatrix}
\hat{a}_{k-1} \\ \hat{a}_{k}^{\prime\prime}
\end{pmatrix}
=
\begin{pmatrix}
1 & 0 \\
0 & i
\end{pmatrix}
\begin{pmatrix}
\sqrt{T_k} & -\sqrt{1-T_k} \\
\sqrt{1-T_k} & \sqrt{T_k}
\end{pmatrix}
\begin{pmatrix}
\hat{a}^{\prime\prime}_{k-1} \\ \hat{a}^\prime_{k}
\end{pmatrix} \nonumber\\
&=
\frac{1}{\sqrt{F_{n-k+3}}}
\begin{pmatrix}
\sqrt{F_{n-k+2}} & -\sqrt{F_{n-k+1}} \\
i\sqrt{F_{n-k+1}} & i\sqrt{F_{n-k+2}}
\end{pmatrix}
\begin{pmatrix}
\hat{a}^{\prime\prime}_{k-1} \\ \hat{a}^\prime_{k}
\end{pmatrix}.
\end{align}
In the setup of Fig.~\ref{fig:Schematic}(c),
this transformation is cascaded from $k=2$ to $k=n$
after the phase rotation $\theta_1=90^\circ$ of the first mode
($\hat{a}_1^{\prime\prime}=i\hat{a}_1^\prime$).
After these transformations, the output annihilation operator of the $k$-th mode is given by
\begin{align}
&\hat{a}_k= \nonumber\\
&\frac{i^k F_{n-k+1}}{\sqrt{F_nF_{n+1}}}\hat{a}^\prime_1
+\sum_{l=2}^{k}\frac{i^{k-l+1}F_{n-k+1}}{\sqrt{F_{n-l+1}F_{n-l+3}}}\hat{a}^\prime_l
-\sqrt{\frac{F_{n-k}}{F_{n-k+2}}}\hat{a}^\prime_{k+1}.
\label{eq:output_1Dcluster}
\end{align}
When all input modes are infinitely $\hat{x}$-squeezed vacuum states ($\hat{x}_k^\prime=0$ for all $k$),
it can be proven from Eq.~(\ref{eq:output_1Dcluster}) that the quadratures of the output modes satisfy
\begin{align}
&\hat{p}_1-\hat{x}_2= 0, \quad \hat{p}_n-\hat{x}_{n-1}= 0, \nonumber \\
&\hat{p}_k-\hat{x}_{k-1}-\hat{x}_{k+1}=0 \quad (2\le k\le n-1),
\end{align}
which are the definition of the $n$-mode linear cluster state in Fig.~\ref{fig:TypesOfEntanglement}(a).
This setting is used for generating the 3-mode cluster state (graph 1) in this experiment.

In this method, the transmissivity $T_k$ approaches a constant value
$(\sqrt5-1)/2$ in the limit of $n\to\infty$.
This means that the linear cluster state is unlimitedly generated
by fixing $T_k=(\sqrt5-1)/2$ for all $k\ge2$, satisfying
\begin{align}
&\hat{p}_1-\hat{x}_2= 0, \quad
&\hat{p}_k-\hat{x}_{k-1}-\hat{x}_{k+1}=0 \quad (k \ge2).
\end{align}
This method is used for generating one-dimensional cluster state in Fig.~\ref{fig:1Dcluster}(a).

\section{Acknowledgements}

This work was partly supported by JST PRESTO (JPMJPR1764),
JSPS KAKENHI (18K14143),
and Nanotechnology Platform Program (Molecule and Material Synthesis) of MEXT, Japan.
S. T. acknowledges Tomonori Toyoda for his support on making electric devices through Nanotechnology Platform Program.
K. T. acknowledges financial support from ALPS.

\section{Author Contributions}

S. T. conceived and planned the project.
S. T. and K. T. designed and constructed the experimental setup, acquired and analyzed the data.
A. F. supervised the experiment.
S. T. wrote the manuscript with assistance from K. T. and A. F.

\section{Competing financial interests}

The authors declare no competing financial interests.


\end{document}